\documentclass[aps,prl,twocolumn,superscriptaddress]{revtex4-2}
\usepackage[T1]{fontenc}
\usepackage[utf8]{inputenc}
\usepackage{amsmath}
\usepackage{graphicx}
\usepackage{pdfpages}
\usepackage{newtxtext,newtxmath}

\makeatletter
\AtBeginDocument{\let\LS@rot\@undefined}
\makeatother

\begin{document}
\title{Thermalizing channel states for rapid qubit heating}
\author{Ziyang You}
\affiliation{Institute of Applied Physics and Materials Engineering, University
of Macau, Macau, China}
\affiliation{School of Electronic, Electrical Engineering and Physics, Fujian University
of Technology, Fuzhou, Fujian, China}
\author{Wenhui Huang}
\affiliation{Shenzhen Institute for Quantum Science and Engineering, Southern University
of Science and Technology, Shenzhen, China}
\affiliation{International Quantum Academy, Shenzhen 518048, China}
\author{Libo Zhang}
\affiliation{Shenzhen Institute for Quantum Science and Engineering, Southern University
of Science and Technology, Shenzhen, China}
\affiliation{International Quantum Academy, Shenzhen 518048, China}
\author{Song Liu}
\affiliation{International Quantum Academy, Shenzhen 518048, China}
\affiliation{Shenzhen Branch, Hefei National Laboratory, Shenzhen 518048, China}
\author{Youpeng Zhong}
\affiliation{International Quantum Academy, Shenzhen 518048, China}
\affiliation{Shenzhen Branch, Hefei National Laboratory, Shenzhen 518048, China}
\author{Yibo Gao}
\affiliation{School of Physics and Optoelectronic Engineering, Beijing University
of Technology, Beijing, China}
\author{Hou Ian}
\email{houian@um.edu.mo}

\affiliation{Institute of Applied Physics and Materials Engineering, University
of Macau, Macau, China}
\begin{abstract}
Although known for negatively impacting the operation of superconducting
qubits, thermal baths are shown to exert qubit control in a positive
way, provided they are properly engineered. We demonstrate an experimental
method to engineer the transduction of microwave driving into heat
flow through a leaky resonator. Given the precise conversion, a qubit
receiving the heat flow obtains a quasi-thermal equilibrium with arbitrary
target temperature in hundreds of nanoseconds. We show that the dynamics
of the quantum transducing process is described by thermalizing channel
states, generated from the double dressings of the resonator by the
semi-classical driving and the qubit-resonator coupling. Their spectrum,
coupling, and driving strength determine the channel rate of energy
flow, along with the relaxation rates of photon leakage into the bath.
The analytical prediction is shown to match well with the experimental
measurements on an Xmon qubit circuit. 
\end{abstract}
\maketitle
\emph{Introduction --} The negative impact that the environment imposes
on the operation of superconducting qubits is well known. Over the
years, uninterrapted efforts are spent on analyzing the sources of
these impacts, including bias noise~\citep{Martinis}, dielectric
loss~\citep{Martinis-1}, 1/f flux noise~\citep{Yoshihara}, and
two-level defects~\citep{McDermott,Schlor,Wang}, not to mention
the many more studies on eliminating them. On the other hand, the
concept of bath or reservoir engineering is introduced to dissipatively
drive quantum systems into desired steady states~\citep{Poyatos,Harrington},
including computationally meaningful ones~\citep{Verstraete}. The
concept has been applied to system specific studies on optical cavities~\citep{Kastoryano},
spin chains~\citep{Morigi}, trapped ions~\citep{Cole-1}, and Rydberg
atoms~\citep{Schonleber} and has been experimentally verified on
superconducting qubits~\citep{Kimchi-Swchartz,Lu,Brown}, as well
as atomic ensembles~\citep{Krauter} and trapped ions~\citep{Lin,Cole,Mourik}.
In particular, when properly engineered, baths play a positive role
in generating entanglement~\citep{Brown,Vollbrecht}. Moreover, properly
engineered baths can facilitate thermalization for various systems~\citep{Yanay,Karimi}
from a thermodynamic perspective.

In fact, quantum heat engines~\citep{Kieu,Quan-1} and subsequently
quantum thermal machines (QTM)~\citep{Tonner,leggio15,bhandari20,Cangemi}
have long been introduced to pursue the microscopic origin of prevalent
concepts in thermo-statistical mechanics. Not only are work and entropy
necessarily reformulated~\citep{Arrachea}, the empirical thermodynamic
laws are also tested~\citep{brandao15,lima-bernado20,Xi} or even
contested~\citep{Lipka} for quantum systems. However, the controllability
of such practical quantum systems as superconducting qubits~\citep{pekola15}
requires faster endothermal and exothermal processes for heat transfer,
demanding specific bath-engineering techniques for QTM. The lack of
such techniques explains the rapid theoretical development of quantum
thermal cycles~\citep{Quan,ian14} including the refinements on Otto
cycles~\citep{karimi16,kloc19,solfan20,Lobejko} and Carnot cycles~\citep{xiao15,kolsloff20,Fadler}
in contrast to the scarcity of their experimental implementations~\citep{Peterson}.

Seeing this need for a precisely controllable non-equilibrium process
for sourcing thermal energy into superconducting qubits, we develop
a technique to transduce microwave driving into heat flux. The energy
conversion is realized by thermalization channel states that are dressed
from a leaky resonator by both the driving and its qubit coupling.
Receiving the heat influx, the target qubit is rapidly saturated to
a steady state on the order of hundred nanoseconds. We analytically
derive the rate of transduction by solving a master equation established
on the channel state spectrum. Experimently, by continuously monitoring
the qubit inversion, we observe the transition of Rabi oscillation
to saturation convergence while the resonator acting as the QTM is
gradually switching on by its driving strength. According to this
driving power as well as the qubit-resonator detuning, the effective
temperature is obtainable over the range from the equilibrating bath
temperature up to infinity. The analytical prediction matches well
with our experimental observations on an Xmon circuit. 

Dynamically speaking, the qubit experiences a transition from an initial
stage of coherent pumping to a stage of heat infusing. The evolution
is therefore a hybridization of stimulated radiation at the beginning
and spontaneous radiation that entails. Since the spontaneous absorption
is sourced from the hybridized driving rather than a higher-temperature
bath, the qubit reaches a hotter thermal state much faster than it
would be with an equilibrating bath. The terminal steady state is
thus a quasi-equilibrium state for the qubit, which balances the influxing
transduced energy and the outfluxing dissipation. In addition, we
note that the recently discovered inverse version~\citep{Shapira}
of quantum Mpemba effect~\citep{Chatterjee,Joshi,Moroder,Zhang}
also enables fast heating of qubits. The qubit states are driven to
higher doublets with a temporally interlaced sequence of coherent
driving and relaxation. In contrast, the thermal photons in the channel
states here are derived from mixing with the bath continuum, thus
limiting the effective temperature to infinity instead of negative
temperatures. A combination of the channel-state thermalization and
inverse Mpemba effect could potentially enable more versatile heating
control to qubits.

\begin{figure}
\includegraphics[bb=85bp 80bp 515bp 370bp,clip,width=8.6cm]{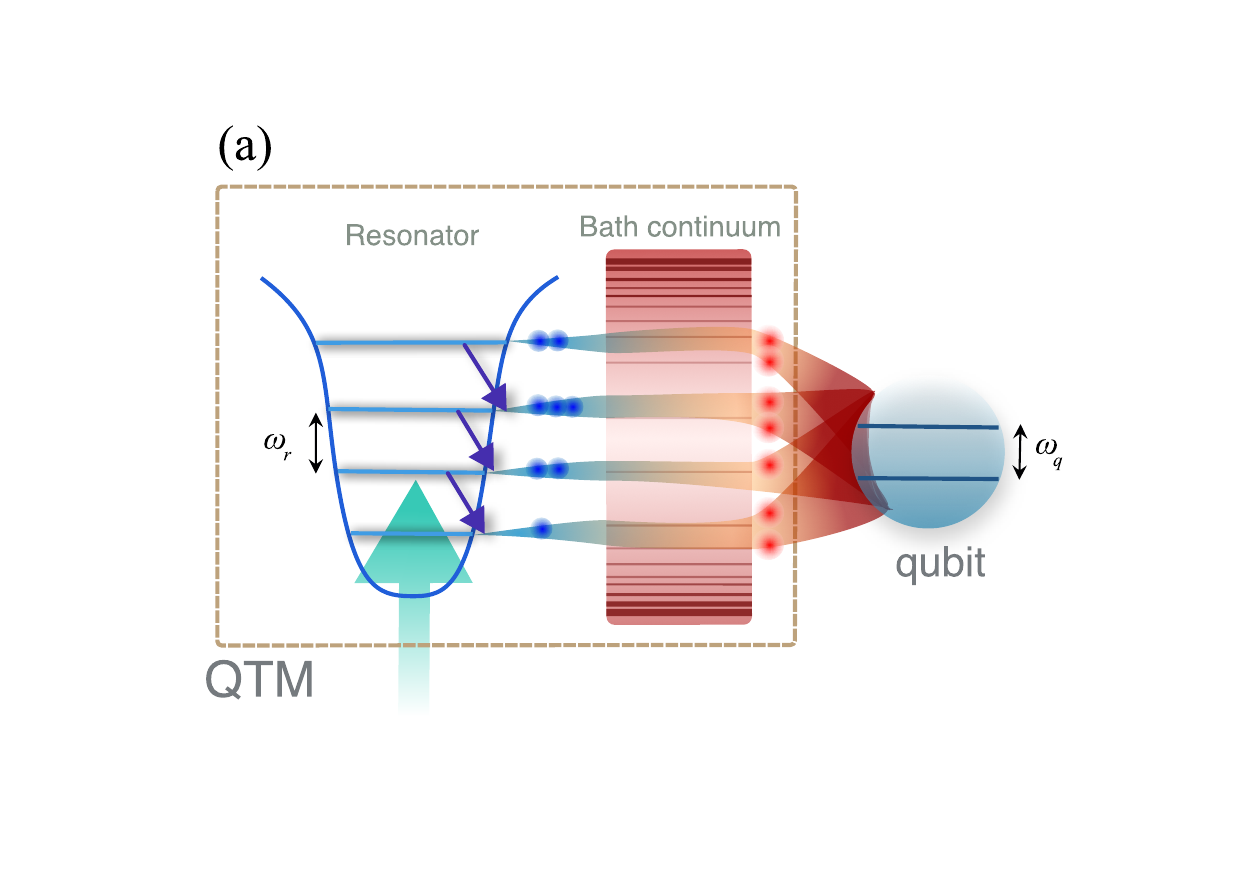}

\includegraphics[bb=40bp 12bp 805bp 300bp,clip,width=8.6cm]{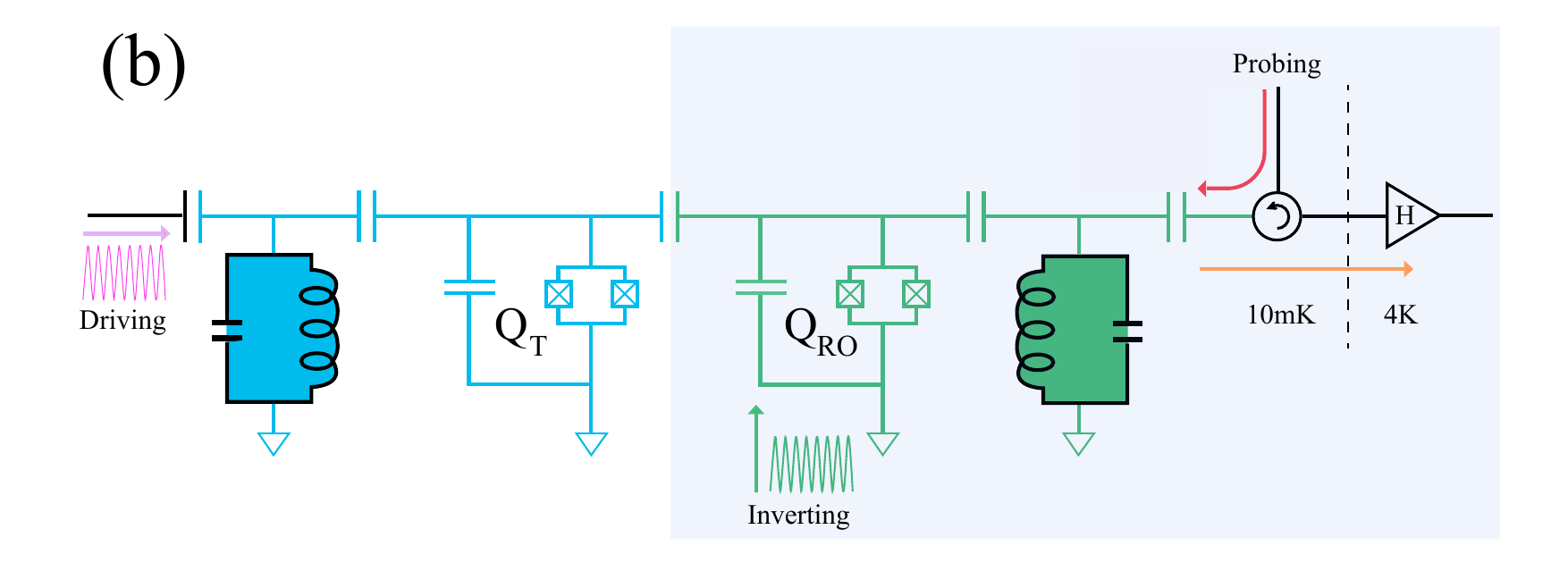}

\caption{(a) Conceptual illustration of thermalizing channels of the QTM constructed
from hybridizing a resonator mode with a continuous multimode bath.
While being energy fed by an external pump, the leaky resonator has
its states broadened by bath couplings so that multi-photon resonance
with the bath are engaged before photons are absorbed by the two-level
qubit. The bath resonance essentially converts single-mode photons
(blue circles) into broad-spectrum thermal photons (red circles) such
that energy carried by the driving photons are rendered into a heat
flow. The LC circuit represents the low-Q coplanar stripped waveguide
that provides the resonator modes for generating the thermalizing
channels. (b) Equivalent circuit on the superconducting qubit chip,
where the target Xmon qubit $Q_{T}$ while being driven from the left
is monitored by a readout subcircuit (the shaded region) on the right.
The readout is realized by a dispersively coupled qubit $Q_{RO}$,
whose frequency shift is registered on another dedicated resonator.~\label{fig:model}}
\end{figure}

\emph{Thermalizing channels --} Transducing an external driving into
a thermalizing energy flow requires a QTM to mediate between the qubit
and the bath, without which the qubit would simply decay and emit
energy. The existence of mediation creates thermalization channels,
which are essentially intermediate levels that accentuate multi-photon
processes across broad spectrum between the qubit and the driving.
As illustrated in Fig.~\ref{fig:model}(a), these intermediate channel
states are formed by a cavity mode dressing the two-level qubit whilst
being pumped by a driving field. The poor quality and thus broad linewidth
of the cavity permits frequent exchanges with the environment and
retainment of photons across a finite band of frequencies, which are
sourced from the monochromatic drive. On the other hand, when the
qubit is biased to near-resonance with the cavity to form photon exchanges
with the cavity via their Jaynes-Cummings coupling. Since the cavity
photons have a continuous spectrum, it is equivalent to a multi-mode
oscillator that models heat baths and induces spontaneous radiation
of the qubit. 

\begin{figure}
\includegraphics[bb=55bp 220bp 540bp 610bp,clip,width=8.6cm]{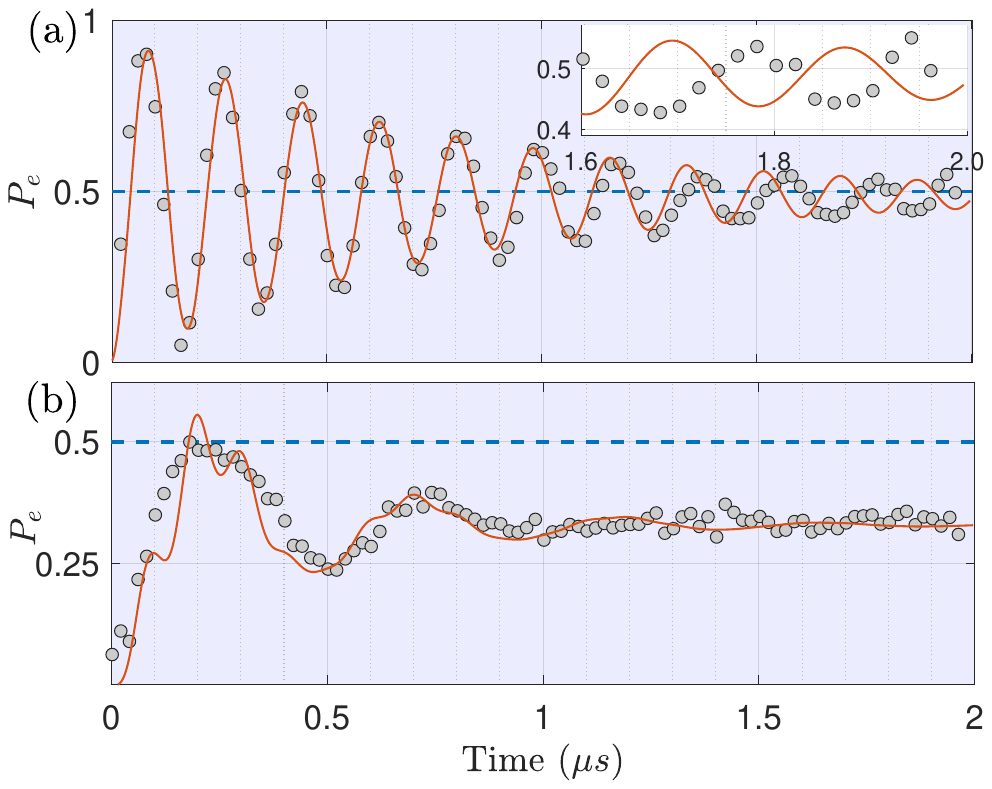}

\caption{Measured excited state population or inversion $P_{e}$ (grey dots)
of the Xmon qubit against time: (a) exhibiting Rabi oscillation (about
the blue $P_{e}=1/2$ dash line) when the driving frequency $\omega_{d}/2\pi=5.46$~GHz
is resonant with the qubit but far detuned from the leaky resonator
frequency $\omega_{r}/2\pi=5.445$~GHz; (b) exhibiting rapid convergence
to a saturated level when $\omega_{d}/2\pi=5.45$~GHz is simultaneously
near resonant with both the qubit and the resonator, which is always
under $P_{e}=1/2$. The orange solid curve are theoretical fit computed
from numerical methods. The inset shows that even when the resonator
is far-detuned from the driving, the bath mediation is effective as
the discrepancy becomes apparent between the measured $P_{e}$ and
the numerical prediction assuming only qubit-drive resonance, as time
increases.~\label{fig:dynamics_types}}
\end{figure}

In our experiment, the target Xmon qubit $Q_{T}$ of level spacing
$\omega_{q}$ is coupled to an $\omega_{r}/2\pi=5.445$~GHz stripline
resonator, whose equivalent circuit is illustrated in Fig.~\ref{fig:model}(b).
The low-Q resonator has a full-width at half-maximum (FWHM) linewidth
of $\gamma_{r}/2\pi1.2$~MHz. An external microwave field of frequency
$\omega_{d}$ is driving it at various strength $\Omega$, from a
few kHz and a few MHz, along the waveguide at d etuning $\Delta_{q}$
with low reflection. When detuned from the resonator but resonant
with the qubit, the microwave field at driving frequency $\omega_{d}/2\pi=\omega_{q}/2\pi=5.46$~GHz
propagates through the resonator and drives the qubit directly, resulting
in the observed Rabi oscillation with a decaying envelope with relaxation
time $T_{1}=5.2\mu$s, as shown in Fig.~\ref{fig:dynamics_types}(a).
When the leaky cavity is closely resonant with the qubit tuned to
$\omega_{q}/2\pi=5.448$~GHz and with a microwave driving frequency
$\omega_{d}/2\pi=5.45$~GHz and strength $\Omega/2\pi=2.0$~MHz,
a thermalized steady state emerges after a short duration of population
oscillation, as shown in Fig.~\ref{fig:dynamics_types}(b) (Cf. Sec.
I and II of~\citep{Supp} for experimental details).

In fact, thermalization behavior already arises in the detuned case
(a) in the form of prolonged Rabi periods. The gradual prolongation
is apparent when we compare the measured data points with the theoretical
expectation, the red curve produced by numerically solving the Liouville-master
equation $\dot{\rho}=\mathcal{L}[\rho]$. The superoperator $\mathcal{L}$
on $\rho$ accounts for both the oscillation dynamics derived from
the waveguide QED Hamiltonian $H=\omega_{q}\sigma_{z}+\Omega(\sigma_{+}e^{-i\omega_{d}t}+\sigma_{-}e^{i\omega_{d}t})$
for the detuned case, where the driving strength is set to $\Omega/2\pi=5.2$~MHz,
and the relaxation dynamics induced by the bath-related Lindbladians.

Comparing the close-resonant scenario with the detuned case, the disappearance
of Rabi oscillation which is substituted by the appearance of a steady
state is explained by the formation of thermalization channels. On
one hand, the resonator-qubit resonance essentially forms dressed
states between these two subsystems. On the other, though the leaky
resonator is constantly pumped by the resonant drive, the short-lived
cavity photons are either exchanged into qubit excitations through
the dressed states or spontaneously radiated into the surrounding
bath. The latter results in thermal photons of differing frequencies
spontaneously fedback from the broad spectral bath, which ultimately
are injected into the qubit. In other words, instead of decayed as
in equilibrating processes with the bath, the spontaneous emission
and absorption are sustained, forming \emph{channel states} that transfer
thermal photon energy sourced from the driving and mediated by the
bath into the qubit, as illustrated in Fig.~\ref{fig:model}(a).

Originating from the tripartite system, the channel states mix basis
states from the $2n$-dim Hilbert spaces of the qubit-resonator subsystem
with the semi-classical driving terms. When they are equilibrating
with a lower-temperature bath, the qubit state within the mixture
approaches a steady state that is in quasi-equilibrium with the bath:
the qubit inversion is saturated to an equivalent Gibbs temperature
higher than the bath temperature, depending on the driving strength.
In the case shown in Fig.~\ref{fig:dynamics_types}(b), the qubit
effective temperature $T$ is $0.37$~K while the bath temperature
set by the dilution refrigerator is about $20$~mK.

\emph{Non-equilibrium dynamics -- }To appreciate the experimental
observations above from a theoretical perspective, we consider the
quasi-equilibrium process generated by the interplay between the channel
states and the bath be described by the master equation $\dot{\rho}=\mathcal{L}[\rho]$
under the driven Jaynes-Cumming (JC)-model Hamiltonian
\begin{equation}
H=\frac{\omega_{q}}{2}\sigma_{z}+\omega_{r}a^{\dagger}a+\eta(a^{\dagger}\sigma_{-}+a\sigma_{+})+\Omega(a^{\dagger}e^{-i\omega_{d}t}+ae^{i\omega_{d}t}).\label{eq:sys_Ham}
\end{equation}
Shown as the solid curve in Fig.~\ref{fig:dynamics_types}(b), the
numerical solution~\citep{Johansson} with a computational basis
of 15 dimensions to the master equation fits well with the experimental
measurements.

To sufficiently describe the energy converted from the coherent driving
of strength $\Omega$ into an incoherent heat flow, we consider the
channel states mentioned above as the eigenstates of Hamiltonian~(\ref{eq:sys_Ham}).
Since the qubit-resonator coupling and the driving terms cannot be
simultaneously diagonalized, approximations have to be sought. Methodology-wise,
a typical circuit QED approach only diagonalize the $\eta$-coupling
term while leaving the weak driving $\Omega$ as a perturbation. In
the opposite case where $\Omega$ is comparatively stronger than $\eta$,
such as the interaction between a superconducting qubit and two-level
oxide defects, the driving is diagonalized in the coherent state space
while the $\eta$ term is regarded as a pertubation~\citep{Wang}.
Here, neither of these two cases applies since $\eta$ and $\Omega$
are on similar orders of magnitude because the leaky resonator, while
being coupled to the qubit, is rapidly driven to sustain the thermalization
against the simultaneous relaxation. Without the perturbation approach,
we use dimension truncation to make the theoretical analysis analytical,
i.e. the number of channel states is limited to three, namely $\{\left|\mu_{k}\right\rangle :k=0,1,2\}$,
which shall prove to be sufficient for interpreting the experimental
results.

These three channel states diagonalize Hamiltonian~(\ref{eq:sys_Ham})
as $H_{\mathrm{S}}=\sum_{k}\varepsilon_{k}\left|\mu_{k}\right\rangle \left\langle \mu_{k}\right|$
in the lowest energy dimensions~\citep{Supp} and are responsible
for converting coherent energy from the drive into thermal energy.
To substantiate this proposition, consider the master equation under
the channel state basis
\begin{multline}
\dot{\rho}=-i\left[H_{\mathrm{S}},\rho\right]-\\
\sum_{n,k\neq n}\Gamma_{kn}\biggl[\frac{1}{2}\left\{ \vert\mu_{n}\rangle\langle\mu_{n}\vert,\rho\right\} -\left|\mu_{n}\right\rangle \left\langle \mu_{k}\right|\rho\left|\mu_{k}\right\rangle \left\langle \mu_{n}\right|\biggr],\label{eq:master_eq}
\end{multline}
which is derived from perturbative analysis to a Liouville equation
on the system and the bath~\citep{Supp}. The coefficients
\begin{multline}
\Gamma_{kn}=\left[\bar{\gamma}(\varepsilon_{n}-\varepsilon_{k})+\bar{\gamma}(\varepsilon_{k}-\varepsilon_{n})+\gamma(\varepsilon_{k}-\varepsilon_{n})\right]\\
\times\frac{\left(\eta(\Delta_{k}^{-1}+\Delta_{n}^{-1})+\Omega(\varepsilon_{k}^{-1}+\varepsilon_{n}^{-1})\right)^{2}}{\left(1+\eta^{2}\Delta_{k}^{-2}+\Omega^{2}\varepsilon_{k}^{-2}\right)\left(1+\eta^{2}\Delta_{n}^{-2}+\Omega^{2}\varepsilon_{n}^{-2}\right)}\label{eq:chan_rate}
\end{multline}
signifies the \emph{channel rate}, in which $\bar{\gamma}(\omega)$
and $\gamma(\omega)$ indicate the contributions of spontaneous transitions
induced by the bath at finite temperature and at vacuum, respectively,
at a given frequency $\omega$. Consequently, $\bar{\gamma}(\varepsilon_{n}-\varepsilon_{k})$
for example indicates the rate of directional (from $k$-th to $n$-th
channel state) spontaneous radiation associated with photon energy
$\varepsilon_{n}-\varepsilon_{k}$ at the bath temperature (20mK of
the dilution fridge). Moreover, as in the usual cases of master equations,
the finite-temperature contribution is two-sided while the vacuum
contribution is one-sided in their frequency distributions with respect
to the bath spectrum. However, different from the usual cases, the
$\Gamma_{kn}$ here is modified by a factor (second line of the equation)
determined by both $\eta$ and $\Omega$ as well as the detunings
$\Delta_{k}=\varepsilon_{k}-(\omega_{q}-\omega_{d})$. At vanishing
coupling strength $\eta$ and weak driving $\Omega$, Eq.(\ref{eq:master_eq})
falls back to the usual master equation of a weakly driven qubit,
whose trajectory would become those shown in Fig.~\ref{fig:dynamics_types}(a).
In addition, if $\Gamma_{kn}$ vanishes by way of reducing $\bar{\gamma}$
and $\gamma$ to vanishing values, the effect of driving would then
embody only in the free evolution generated by $H_{\mathrm{S}}$,
i.e. Rabi oscillations.

\begin{figure}
\includegraphics[clip,width=8.6cm]{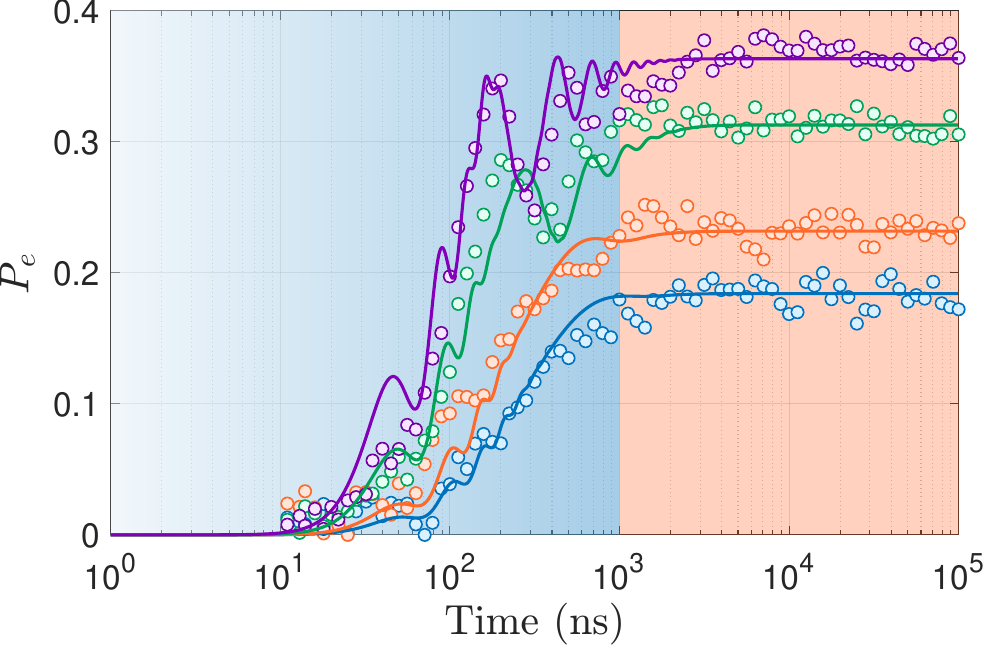}

\caption{Long-term population $P_{e}$ variation of the qubit excited state
on a semi-log scale over time at driving strengthes $\Omega/2\pi=1.5$~MHz
(blue), 2~MHz (orange), 3.5~MHz (green), and 5~MHz (purple). The
circled dots are the experimental measurements. The solid curves are
theoretical predictions from the analytical solutions to the master
equation using the same experimental $\Omega$ values as parameters,
showing a great fit to the measurements. The background is shaded
in two colors to visually discern the separation of two dynamic stages:
(i) the light-blue endothermic stage and (ii) the light-orange quasi-equilibrium
stage.~\label{fig:dynam_vs_driving}}
\end{figure}

\emph{Thermal photon transduction} -- On the contrary, with finite
coupling and sufficient driving, $\Gamma_{kn}$ being the coefficient
of the Lindbladian describes the channeling rate of energy towards
the qubit. The factor of $\bar{\gamma}$ and $\gamma$, on one hand,
demonstrates the necessary presense of the bath to substantiate broadband
photon exchanges; the factor of $\eta$ and $\Omega$, on the other,
determines the conversion from coherent driving photons to spontanesouly
radiated photons. When the driving photons following the circuit waveguide
are incident on the resonator, some propagate through the resonator
to directly invert the qubit, whereas the other leak into the bath
and then are reabsorbed by the resonator before entering the qubit
through resonator-qubit resonance. The time difference between these
two processes causes a delay that separates the evolution dynamics
into two stages from a thermodynamic perspective: (i) an endothermic
stage and (ii) a quasi-equilibrium stage. As shown in Fig.~\ref{fig:dynam_vs_driving}
on a log-time scale, this two-stage distinction in dynamics is uniform
across all driving strengthes where $\Omega/2\pi$ is 1.5, 2, 3.5,
and 5~MHz for the four curves from bottom to top (See Sec.~IV of~\citep{Supp}
for the analytic derivations).

Starting from a bath-equilibrating ground state (due to the low environmental
temperature), the qubit absorbs driving photons mediated by the resonator
QTM. If the resonator has a high Q-factor as in typical circuit QED
systems, qubit inversion would persist until the excited state population
$P_{e}$ approaches near unity; but the low-Q resonator permits rapid
photon leakage into the bath, leaving room for photons to be refilled
by the qubit. Hence, oscillations with local peaks much less than
unity appear. Meanwhile, the leaked photons are fedback from the broad-spectrum
bath as incoherent thermal photons and reabsorbed by the resonator.
They would channel into the qubit via the channel states and give
rise to an increasing mean that carries the coherent oscillation.
These two processes, expressed by the two terms on the RHS of Eq.~(\ref{eq:master_eq}),
compete with each other~\citep{Wang} and their concurrency effects
the endorthermic stage.

Later, when the population $P_{e}$ accumulates, the upward and downward
transition rates between the qubit states begin to reach a detailed
balance and $P_{e}$ would saturate at a steady value dependent on
the driving strength $\Omega$. Thermodynamically, it is the balance
between the energy influx due to the driving and the energy outflux
towards the bath, rather than a balance with the bath only. The stage
appearing as a steady state is thus a quasi-equilibrium state: the
qubit system remains out of equilibrium with the low-temperature bath
throughout. For instance, while the bath remains at 20~mK, the four
steady states in Fig.~\ref{fig:dynam_vs_driving} correspond to effective
temperatures 150~mK, 190~mK, 300~mK, and 450~mK, respectively,
if we assume Gibbs distribution.

These experimental observations coincide with the theoretical results
predicated by Eq.~(\ref{eq:master_eq}), which is exactly solvable.
The non-diagonal elements $\rho_{jk}$ ($j\neq k$) are decoupled
and the first-order equation leads to a solution with negative real
and imaginary exponentials of significant magnitudes. It signifies
rapidly decaying transitions along with overriding oscillations among
the channel states, which translates into a rapidly converging envelope
on the order of $\mu$s for the bare qubit state. Fast heating therefore
appears during the endothermic stage, as we observe in Fig.~\ref{fig:dynam_vs_driving}.
Simultaneously, the diagonal elements $\rho_{kk}$ are coupled but
still solvable by diagonalizing the transition matrix. The solution
$\rho_{kk}(t)$, containing real exponentials $-\sum_{n,k\neq n}\Gamma_{kn}$,
verifies the observation about rapid convergence above and predicts
the steady state of the qubit inversion,
\begin{equation}
P_{e,SS}=\frac{1}{3}\sum_{k}\frac{\Omega^{2}}{\Omega^{2}+\varepsilon_{k}^{2}(1+\eta^{2}/\Delta_{k}^{2})}\label{eq:steady_sta}
\end{equation}
that is observed during the quasi-equilibrium stage. It becomes clear
from Eq.~(\ref{eq:steady_sta}) that the inversion at steady state
in general increases with the driving strength $\Omega$.

\begin{figure}
\includegraphics[clip,width=8.6cm]{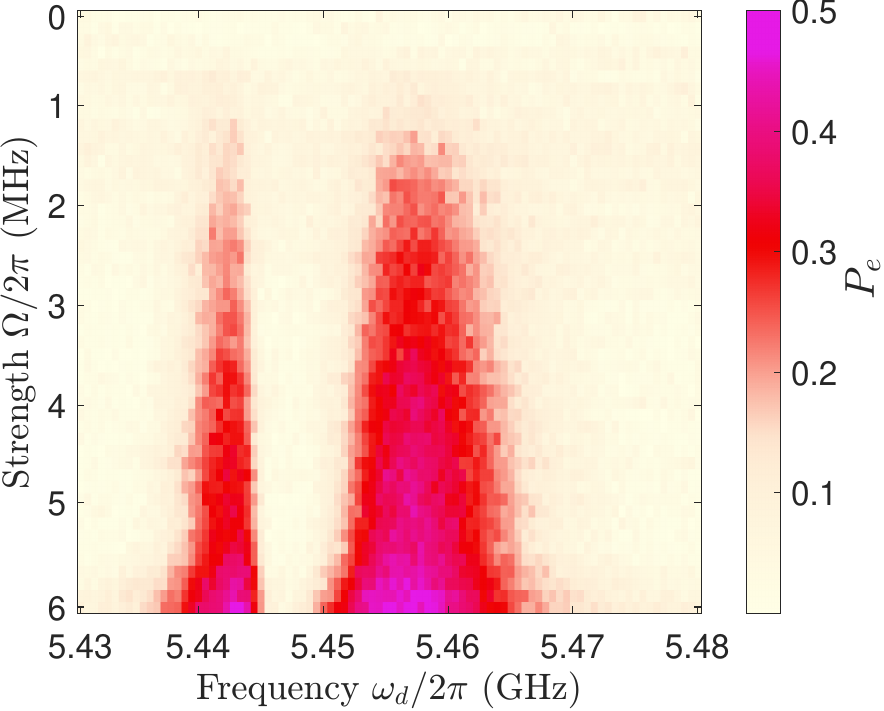}

\caption{Qubit inversion $P_{e}$ plotted (a) as contour level against the
driving frequency $\omega_{d}$ and driving strength $\Omega$; and
(b) against $\omega_{d}$ only at the fixed strength $\Omega/2\pi=5.8$~MHz.~\label{fig:chan_width}}
\end{figure}

To be more specific, we have obtained the steady-state inversion at
various driving frequencies and driving strengthes, which are plotted
in Fig.~\ref{fig:chan_width}. Effective temperatures are obtainable
full range from near 0 to $\infty$ over the increase of $\Omega$.
In addition, enhancing qubit inversion is particularly conspicuous
when the driving frequency is either $\omega_{d}/2\pi=5.442$ or 5.457~GHz,
which is neither resonant with the resonator nor the qubit. Rather,
it conforms to the expectation of Eq.~\ref{eq:chan_rate} that the
channeling rates are maximized when the channel states frequencies
$\varepsilon_{k}$ and their detunings $\Delta_{k}$ from the relative
qubit frequency $\delta_{q}$ are simultaneously optimized. Given
the two factors in the equation, $\varepsilon_{k}$ would need to
match, on one hand, the bath spectrum to maximize the spontaneous
resonator-bath photon exchange and atone, on the other hand, to $\eta$
such that the exchanged photons are synchronously reabsorbed by the
qubit.

In summary, we introduce a method of transduction to convert coherent
driving into heat infusion on superconducting qubits. The experiment
is conducted using a leaky and thus strongly bath-coupled resonator
as an interface between the driving and a target qubit. Depending
on the driving parameters, full range of effective temperatures is
observed at quasi-equilibrium after 100s of nanosecond. The rapid
heating is theoretically explained by the existence of channel states
identified as the diagonalizing states of the tripartite system comprising
the qubit, the resonator, and the driving. They serve as the mediator
with the bath to convert single-mode photons into thermal broadband
photons, resulting in a channel transduction rate determined by the
strength and detuning of the driving and the bath spectrum. The method
permits simplified implementations of quantum thermal processes and
cycle on superconducting qubits.
\begin{acknowledgments}
Y.-P. Z. thanks the support of NSFC of China under grant 12174178
and the Science, Technology and Innovation Commission of Shenzhen
Municipality under grant KQTD20210811090049034. H. I. thanks the support
of FDCT of Macau under grants 0179/2023/RIA3.
\end{acknowledgments}

\newpage
		


\section*{Supplementary material}

# Supplementary Materials to
# Thermalizing channel states for rapid qubit heating

Ziyang You,[1] Wenhui Huang,[2,3] Libo Zhang,[4,3] Song
Liu,[2,3] Youpeng Zhong,[2,3] Yibo Gao,[5] and Hou Ian[1,*]

[1]Institute of Applied Physics and Materials Engineering, University of Macau, Macau, China
[2]Shenzhen Institute for Quantum Science and Engineering,
Southern University of Science and Technology, Shenzhen, China
[3]International Quantum Academy, Shenzhen 518048, China
[4]Shenzhen Institute for Quantum Science and Engineerin g,
Southern University of Science and Technology, Shenzhen, China
[5]School of Physics and Optoelectronic Engineering,
Beijing University of Technology, Beijing, China

## I. EXPERIMENTAL SETUP

The experimental setup is established on a general purposed superconducting circuit with two effective Xmon qubits [1] working in serial connection, as depicted in the circuit diagram in Fig. 1. Each qubit is coupled to a cavity. The target qubit, denoted as $Q_T$, is connected to a fast-decay cavity that constitutes the main part of the QTM model introduced in the main text. The second qubit $Q_{RO}$ is responsible for detecting $Q_T$ and is dispersively coupled to a readout cavity. This readout cavity is connected to a probing signal via a Purcell filter, which suppresses resonator-mediated qubit decay [2]. The probing response signal, reflected from the readout resonator, is routed through a circulator for

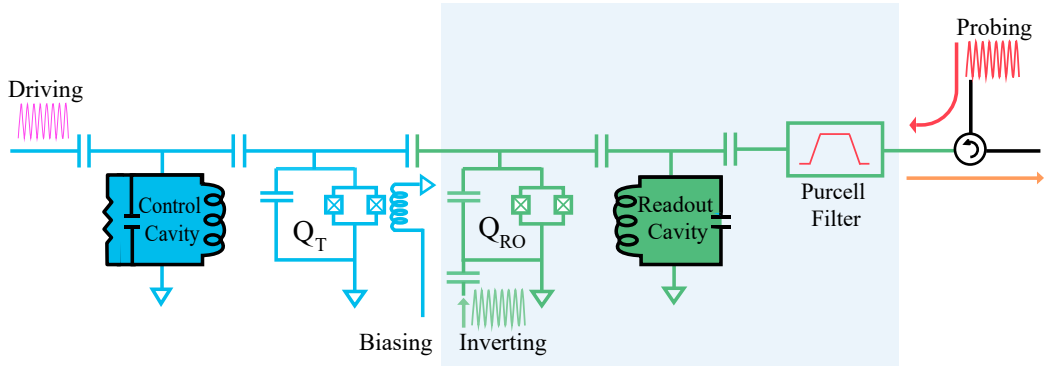

Figure 1. The circuit diagram of the testing chip. A specialized readout protocol is employed, where $Q_{RO}$ serves as an ancillary prober to read out the state of the test qubit. The components in the entire shaded region function as an equivalent readout cavity coupled to the target qubit $Q_T$.

---

* houian@um.edu.mo

| System parameters | Target qubit $Q_T$ | Ancillary qubit $Q_{RO}$ |
|---|---|---|
| Qubit frequency | 5.46 GHz | 3.81 GHz |
| Frequency of coupling resonator | 5.445 GHz | 5.32 GHz |
| Qubit-resonator coupling strength | 4 MHz | 32 MHz |
| Qubit-qubit coupling strength | 100 MHz | |
| Qubit relaxation time | 5.2 $\mu$s | 26 $\mu$s |
| Resonator linewidth | 1.2 MHz | 1.3 MHz |

Table I. Parameters measured on the target and the ancillary qubits, along with their coupling resonators. Under the setting for forming the thermalizing channels, target qubit $Q_T$ is near resonant with a control resonator.

analysis. To perform state tomography on $Q_T$, we employ a specialized readout protocol, using all three components in the shaded region equivalently as a conventional readout resonator.

The circuit chip is fabricated on a c-plane sapphire substrate with an approximately 100 nm thick aluminum layer, deposited using a high-vacuum e-beam evaporator system. The circuit geometries, excluding the junctions, are patterned through photolithography and etched using an inductively coupled plasma (ICP) dry etcher with BCl3/Cl2. The Josephson junctions are patterned via e-beam lithography and fabricated through double-angle evaporation of aluminum followed by liftoff. The measured parameters of the qubits and the resonators are summarized in Tab. I.

## II. READOUT METHODOLOGY

In systems where the qubit has a dedicated readout resonator, the control and readout pulse sequences include a probing pulse applied to the resonator after qubit manipulation. The response to this probing pulse detects the state-dependent frequency shift in the resonator, revealing the qubit state. However, in our circuit fashioned from a general purposed sample chip, target qubit $Q_T$ lacks a dedicated readout resonator to directly measure this state-dependent shift. Instead, a modified readout technique is employed, where the state of $Q_T$ is mapped onto a $Q_{RO}$ via their inter-qubit dispersive coupling, rather than relying on a resonator [3].

As shown in Fig. 1, an additional inverting pulse is applied to the ancillary qubit $Q_{RO}$ after the manipulation of $Q_T$. It is a $\pi$ pulse that fully inverts $Q_{RO}$ when $Q_T$ stays in the ground state. The population of $Q_{RO}$ is subsequently read out by its dedicated readout resonator with high fidelity. The final population of $Q_{RO}$ changes under the same inverting pulse due to the difference in absorption, depending on the state of $Q_T$. As a result, $Q_T$ state can be inferred with good contrast through the $Q_{RO}$ population using this approach, eliminating the need for a dedicated readout resonator.

Details of the apparatus setup are shown in Fig. 2. The sample chip is placed in an aluminum box and mounted in a dilution refrigerator, which maintains a base temperature of approximately 20 mK during measurements. Control and readout pulses for the testing system are generated using IQ modulation. These pulses are shaped in IQ mixers by combining a local oscillator (LO) signal with an envelope generated by arbitrary waveform generators (AWG). A high-precision DC source is used for local flux control of the qubit through the Z-bias line.

After attenuation, the modulated signal is sent to the testing system. The readout response from the resonator is first amplified by a high electron mobility transistor (HEMT) amplifier at the 4 K stage, followed by a low-noise amplifier (LNA) at room temperature. The amplified signal is then

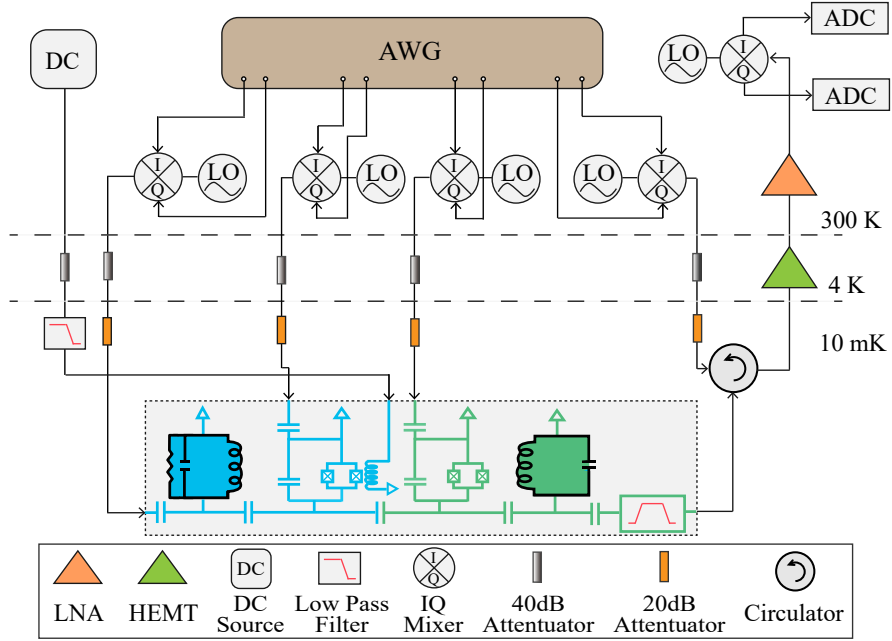

Figure 2. The signal chain of the experiment involves generating control and readout pulses using IQ modulation, combining a local oscillator (LO) signal with an envelope from an arbitrary waveform generator (AWG). The readout response is amplified by a high electron mobility transistor (HEMT) at 4 K and a low-noise amplifier (LNA) at room temperature, then demodulated in an IQ mixer and recorded by analog-to-digital converters (ADCs).

demodulated in an IQ mixer and recorded by analog-to-digital converters (ADCs).

## III. CONSTRUCTING THE MASTER EQUATION

First, let the Hamiltonian in Eq. (1) of the main text be written as

$$H' = \frac{\omega_q}{2}\sigma_z + \omega_0 a^\dagger a + \eta(a\sigma_+ + a^\dagger\sigma_-) + \Omega(a^\dagger + a) \tag{1}$$

in the rotating frame of the driving $V = \omega_d a^\dagger a + \omega_d \sigma_z$. If we truncate the Hilbert space dimension of the bare states to the lowest three, i.e. $\{|0,g\rangle, |1,g\rangle, |0,e\rangle\}$, and let the ground reference of the qubit be offset by $\omega_q/2$, the effective Hamiltonian is represented by the 3x3 matrix

$$H' \approx \begin{pmatrix} \delta_q & \eta & 0 \\ \eta & \delta_0 & \Omega \\ 0 & \Omega & 0 \end{pmatrix} \tag{2}$$

where the detunings are defined as $\delta_q = \omega_q - \omega_d$ and $\delta_0 = \omega_0 - \omega_d$. At close qubit-cavity resonance with $\delta_0 \approx \delta_q$, the eigenenergies of $H'$ are the roots of the cubic equation

$$\varepsilon^3 - 2\delta_q \varepsilon^2 - (\Omega^2 + \eta^2 - \delta_q^2)\varepsilon + \Omega^2 \delta_q = 0. \tag{3}$$

The associated depressed cubic is

$$\lambda^3 - \left(\Omega^2 + \eta^2 + \frac{\delta_q^2}{3}\right)\lambda + \frac{\delta_q}{3}\left(\Omega^2 - 2\eta^2 + \frac{2\delta_q^2}{9}\right) = 0 \tag{4}$$

where $\lambda = \varepsilon - 2\delta_q/3$, giving the cubic root discriminant

$$\mathscr{D} = 4\eta^2 \delta_q^4 + (\Omega^4 + 20\Omega^2\eta^2 - 8\eta^4)\delta_q^2 + 4(\Omega^2 + \eta^2)^3. \tag{5}$$

Writing the $\delta_q^2$ coefficient as $(\Omega^2 + (10 + 6\sqrt{3})\eta^2)(\Omega^2 + (10 - 6\sqrt{3})\eta^2)$, we recognize that as long as $\Omega > \sqrt{6\sqrt{3} - 10}\,\eta$ the discriminant is guaranteed positive, making Eq. 4 admit three real roots. Since the driving strength $\Omega$ and the coupling strength $\eta$ is physically on the same order of magnitude, this inequality is easily satisfied; if not, $\mathscr{D}$ would still have three real roots, where the inequality difference would be offset by the other positive terms. Therefore, we can solve the cubic equation by Viète's method, giving the roots

$$\varepsilon_k = \frac{2\delta_q}{3} + \frac{2\xi}{\sqrt{3}}\cos\frac{1}{3}\left[\cos^{-1}\frac{\sqrt{3}\delta_q(\delta_q^2/9 + 3\eta^2 - \xi^2)}{2\xi^3} + 2k\pi\right] \tag{6}$$

where $k$ runs through $\{0, 1, 2\}$ and $\xi^2 = \Omega^2 + \eta^2 + \delta_q^2/3$. Since $\varepsilon_k$ are distinct, they have unique eigenvectors

$$|\mu_k\rangle = \frac{\eta\varepsilon_k\,|0, e\rangle + \Delta_k\varepsilon_k\,|1, g\rangle + \Omega\Delta_k\,|0, g\rangle}{\sqrt{\eta^2\varepsilon_k^2 + \Delta_k^2\varepsilon_k^2 + \Omega^2\Delta_k^2}} \tag{7}$$

where $\Delta_k = \varepsilon_k - \delta_q$. In other words, Eq. (7) forming the new eigenbasis are the channel states that absorb driving energy and emit thermal energy, on which the system Hamiltonian is written as $H_S = \sum_k \varepsilon_k\,|\mu_k\rangle\langle\mu_k|$. The annihilation operator is consequently expanded as

$$a = \sum_{j,k} \nu_{jk}\,|\mu_j\rangle\langle\mu_k| \tag{8}$$

and similarly for the creation operator, where

$$\nu_{jk} = \frac{\Omega\varepsilon_k}{\sqrt{\Omega^2 + \varepsilon_j^2(1 + \eta^2/\Delta_j^2)}\sqrt{\Omega^2 + \varepsilon_k^2(1 + \eta^2/\Delta_k^2)}}. \tag{9}$$

The master equation is obtained under the new basis by expanding the Liouville equation to the second perturbation order and tracing over the bath variables, i.e. for a universe density matrix $\rho(t) \otimes w$ at time $t$,

$$\frac{d\rho(t)}{dt} = -\int_0^t d\tau \operatorname{tr}_B\left\{[\mathcal{H}(t), [\mathcal{H}(\tau), \rho(\tau) \otimes w]]\right\} \tag{10}$$

where $\mathcal{H} = \sum_j \kappa_j (a + a^\dagger)(b_j + b_j^\dagger)$ is the cavity-bath coupling Hamiltonian [4]. The coefficients $\kappa_j$ determine the spectrum density of the bath and they contribute to the frequency dependent relaxation rate $\gamma(\omega) = 2\pi \sum_j \kappa_j^2 \delta(\omega - \omega_j)$ when assuming a typical Ohmic bath spectrum. During the system evolution, the bath being an energy reservoir remains stable and unentangled with the system such that the coupling $\mathcal{H}(t) = \sum_j \kappa_j (a(t) + a^\dagger(t))(b_j(t) + b_j^\dagger(t))$ at time $t$ has the system and the bath operators remain orthogonal, i.e.

$$a(t) = e^{-iH_S t} a e^{iH_S t} = \sum_{j,k} \nu_{jk} e^{-i(\varepsilon_j - \varepsilon_k)t} |\mu_j\rangle \langle \mu_k| \tag{11}$$

$$b_j(t) = b_j e^{-i\omega_j t} \tag{12}$$

With the customary Born-Markov approximation by assuming $\rho(t)$ a slow variable, the double commutator of Eq. (10) is traced to four integrable terms, with tracable factors reducing to

$$\text{tr}_B \left\{ \left( b_j^\dagger b_j e^{i\omega_j(t-\tau)} + b_j b_j^\dagger e^{-i\omega_j(t-\tau)} + (b_j^2 e^{-i\omega_j(t+\tau)} + \text{h.c.}) \right) w \right\} = \bar{n}_j e^{i\omega_j(t-\tau)} + (\bar{n}_j + 1) e^{-i\omega_j(t-\tau)} \tag{13}$$

where inter-bath-mode correlations are ignored and $\bar{n}_j$ denotes the average number of $j$-th mode thermal photons. To simplify the integration, we reverse the integration direction by letting $\tau \to (t-\tau)$ in Eq. (10) such that $(t-\tau) \to \tau$ and the integration range $\int_0^t d\tau$ remains unchanged. Then, expanding the commutators and taking the long-time limit $t \to \infty$, the first term reads

$$\sum_j \kappa_j^2 \sum_{lkn} (\nu_{lk} + \nu_{kl})(\nu_{kn} + \nu_{nk}) e^{-i(\varepsilon_l - \varepsilon_k)t}$$
$$\int_0^\infty d\tau \left[ \bar{n}_j e^{i\omega_j \tau} + (\bar{n}_j + 1) e^{-i\omega_j \tau} \right] e^{-i(\varepsilon_k - \varepsilon_n)(t-\tau)} |\mu_l\rangle \langle \mu_n| \rho(t). \tag{14}$$

The exponentials are integrable via the usual analytic continuation of $\tau$ into the complex domain, where the integration path is closed via a semicircular loop with poles on the imaginary axis. The integral produces residues of delta functions such that Eq. (14) becomes

$$\sum_{lkn} (\nu_{lk} + \nu_{kl})(\nu_{kn} + \nu_{nk}) e^{-i(\varepsilon_l - \varepsilon_n)t} \sum_j \pi \kappa_j^2 \left[ \bar{n}_j \delta(\omega_j + \varepsilon_n - \varepsilon_k) + (\bar{n}_j + 1) \delta(\omega_j - \varepsilon_k + \varepsilon_n) \right] |\mu_l\rangle \langle \mu_n| \rho(t). \tag{15}$$

According to secular approximation [5, 6], the terms involving the discrete levels $\varepsilon_l \neq \varepsilon_n$ do not contribute to the bath couplings to the system modes. Thus, assigning respectively the finite- and the zero-temperature relaxation rates

$$\bar{\gamma}(\omega) = 2\pi \sum_j \kappa_j^2 \bar{n}_j \delta(\omega_j - \omega), \tag{16}$$

$$\gamma(\omega) = 2\pi \sum_j \kappa_j^2 \delta(\omega_j - \omega), \tag{17}$$

Eq. (14) eventually becomes

$$\sum_{k,n} \frac{\Gamma_{kn}}{2} |\mu_n\rangle \langle \mu_n| \rho(t). \tag{18}$$

The coefficients associated with the $|\mu_n\rangle$ term

$$\Gamma_{kn} = \frac{\Omega^2 (\varepsilon_k + \varepsilon_n)^2}{(\Omega^2 + \varepsilon_k^2 (1 + \eta^2/\Delta_k^2)) \, (\Omega^2 + \varepsilon_n^2 (1 + \eta^2/\Delta_n^2))} \, [\bar{\gamma}(|\varepsilon_k - \varepsilon_n|) + \gamma (\varepsilon_k - \varepsilon_n)] \,, \tag{19}$$

combine the coupling strength $\eta$ and the driving strength $\Omega$ with the relaxation rates $\bar{\gamma}$ and $\gamma$. They can be recognized as the energy channeling rates between the $n$-th and $k$-th channel states. In particular, given the definition of the relaxation rates, the zero and negative frequency rates vanish, i.e. $\gamma(\omega) = \bar{\gamma}(\omega) = 0$ whenever $\omega \leq 0$, for any quantum heat bath such as the ohmic bath . That means $\Gamma_{kk} = 0$ and either $\bar{\gamma}(\varepsilon_k - \varepsilon_n)$ or $\bar{\gamma}(\varepsilon_n - \varepsilon_k)$ survives for finite-temperature relaxation, thus the argument $|\varepsilon_k - \varepsilon_n|$ in Eq. (19). In other words, the contributions by finite-temperature relaxations are symmetric about the exchange of $n$ and $k$, while that by zero-temperature relaxation are not.

The other three terms in the integral of Eq. (10) can be found using similar routines, giving rise to the master equation in Lindblad form

$$\dot{\rho} = -i \, [H_{\mathrm{S}}, \rho] - \sum_n \left[ \frac{\Gamma_n^\circ}{2} \{ |\mu_n\rangle \langle \mu_n|, \rho \} - \sum_{k \neq n} \Gamma_{kn} \, |\mu_n\rangle \, \langle \mu_k| \, \rho \, |\mu_k\rangle \, \langle \mu_n| \right] \tag{20}$$

where $\{\cdot, \cdot\}$ denotes the anti-commutator and $\Gamma_n^\circ = \sum_k' \Gamma_{nk}$ (the summation skips the $k = n$ term) the total relaxation rate from the $n$-th channel state.

## IV. SOLVING THE MASTER EQUATION

The master Eq. (20) can be solved by treating each element of the $3 \times 3$ density matrix $\rho(t)$ individually and solving its associated ordinary differential equation. That is, breaking up the master equation in its matrix elements, we have

$$\frac{d\rho_{jk}}{dt} = -i(\varepsilon_j - \varepsilon_k)\rho_{jk} - \left[ \frac{\Gamma_j^\circ + \Gamma_k^\circ}{2} \rho_{jk} - \sum_{m \neq j} \Gamma_{mj} \delta_{jk} \rho_{mm} \right]. \tag{21}$$

Thus, for the non-diagonal elements ($j \neq k$) and diagonal elements ($j = k$), we have respectively

$$\frac{d\rho_{jk}}{dt} = - \left[ i(\varepsilon_j - \varepsilon_k) + \frac{\Gamma_j^\circ + \Gamma_k^\circ}{2} \right] \rho_{jk}, \tag{22}$$

$$\frac{d\rho_{jj}}{dt} = -\Gamma_j^\circ \rho_{jj} + \sum_{m \neq j} \Gamma_{mj} \rho_{mm}. \tag{23}$$

Equation (22) can be directly inverted to find

$$\rho_{jk}(t) = \rho_{jk}(0) \exp \left\{ -i(\varepsilon_j - \varepsilon_k) - \frac{\Gamma_j^\circ + \Gamma_k^\circ}{2} \right\} t \tag{24}$$

while Eq. (23) is also invertible by regarding it as the matrix differential equation $\dot{\boldsymbol{\rho}} = G\boldsymbol{\rho}$ where $\boldsymbol{\rho} = (\rho_{00}, \rho_{11}, \rho_{22})$ is the vector of the diagonal elements and

$$G = \begin{pmatrix} -\Gamma_0^\circ & \Gamma_{10} & \Gamma_{20} \\ \Gamma_{01} & -\Gamma_1^\circ & \Gamma_{21} \\ \Gamma_{02} & \Gamma_{12} & -\Gamma_2^\circ \end{pmatrix}. \tag{25}$$

The individual diagonal elements are solvable by diagonalizing $G$ as $P^{-1}\Lambda P$. The corresponding eigenvalues, i.e. the diagonal elements of $\Lambda$, are the roots of

$$\lambda^3 + p\lambda^2 + q\lambda = 0 \tag{26}$$

where the coefficients are

$$p = \Gamma_0^\circ + \Gamma_1^\circ + \Gamma_2^\circ, \tag{27}$$
$$q = \Gamma_0^\circ\Gamma_1^\circ + \Gamma_1^\circ\Gamma_2^\circ + \Gamma_2^\circ\Gamma_0^\circ - \Gamma_{01}\Gamma_{10} - \Gamma_{12}\Gamma_{21} - \Gamma_{20}\Gamma_{02}. \tag{28}$$

Hence, besides $\lambda_0 = 0$, the non-zero roots are

$$\lambda_\pm = -\frac{1}{2}\sum_j \Gamma_j^\circ \pm \frac{1}{2}\sqrt{\sum_j \left[(\Gamma_j^\circ - \Gamma_{j^+}^\circ)^2 + 4\Gamma_{jj^+}\Gamma_{j^+j} - \Gamma_{j^+}^{\circ 2}\right]} \tag{29}$$

where the index $j$ is meant to iterate over $\{0, 1, 2\}$ and $j^\pm$ denotes the modular increment or decrement, i.e. $(j \pm 1 \mod 3)$.

The analytical solution to Eq. (23) is then

$$\boldsymbol{\rho}(t) = P^{-1}e^{\Lambda t}P\boldsymbol{\rho}(0). \tag{30}$$

Considering that all the $\Gamma$ are positive and the magnitude of $4\Gamma_{jj^+}\Gamma_{j^+j}$ is close to $\Gamma_{j^+}^{\circ 2} = (\Gamma_{jj^+} + \Gamma_{j^-j^+})^2$, the discriminant of Eq. (29) is either positive (in most cases) or a small negative value (in rare cases). Even when the latter occurs, the dominant magnitude of $\lambda_\pm$ is contributed by the real part of the $\Gamma_j^\circ$ sum. This shows the time evolutions of the diagonal elements given by Eq. (30) will mostly retain a decaying rather than oscillating form. On the other hand, given the relatively large magnitude of $(\varepsilon_j - \varepsilon_k)$, the non-diagonal elements given by Eq. (24) will be oscillating under the decaying envelopes.

## V. QUASI-EQUILIBRIUM STATE

To obtain the steady state corresponding to the quasi-equilibrium limit, we can simply observe from Eq. (22) that the non-diagonal elements would approach zero as expected and from Eq. (23) that the steady state is just the non-trivial solution to the matrix equation associated with the $G$ matrix in Eq. (25). The existence of the eigenvalue $\lambda_0 = 0$ shows that $G$ is singular and the non-trivial solution is exactly the eigenvector corresponding to $\lambda_0$, namely

$$\boldsymbol{\rho} = \frac{1}{Z}\begin{bmatrix} \Gamma_1^\circ\Gamma_2^\circ - \Gamma_{12}\Gamma_{21} \\ \Gamma_2^\circ\Gamma_0^\circ - \Gamma_{20}\Gamma_{02} \\ \Gamma_0^\circ\Gamma_1^\circ - \Gamma_{10}\Gamma_{01} \end{bmatrix} \tag{31}$$

where $Z$ denotes the normalization constant. Using the notation introduced in the last section, the vector elements can be summarized as

$$\rho_{kk} = \frac{\Gamma_{k^+}^\circ \Gamma_{k^-}^\circ - \Gamma_{k^+ k^-} \Gamma_{k^- k^+}}{\sqrt{\sum_j \left[ \Gamma_{j^+}^\circ \Gamma_{j^-}^\circ - \Gamma_{j^+ j^-} \Gamma_{j^- j^+} \right]^2}}. \tag{32}$$

These diagonal elements are still in the channel state $\{|\mu_k\rangle\}$ basis. Using Eq. (7) and tracing out the cavity space, the excited state population can be found as

$$\begin{aligned}
P_{e,SS} &= \sum_k \rho_{kk}^2 \, |\langle e, 0 | \mu_k \rangle|^2 \\
&= \sum_k \frac{\eta^2 (\Gamma_{k^+}^\circ \Gamma_{k^-}^\circ - \Gamma_{k^+ k^-} \Gamma_{k^- k^+})^2}{\eta^2 + \Delta_k^2 (1 + \Omega^2/\varepsilon_k^2)} \Bigg/ \sum_j \left[ \Gamma_{j^+}^\circ \Gamma_{j^-}^\circ - \Gamma_{j^+ j^-} \Gamma_{j^- j^+} \right]^2.
\end{aligned} \tag{33}$$



\end{document}